\documentclass[twocolumn,aps,showpacs,showkeys]{revtex4}
\usepackage{epsfig,amsmath,amssymb}
\usepackage{color}

\begin{document}

\title{Spin-1/2 $XXZ$ diamond chain within the Jordan-Wigner fermionization approach}
\author{T. Verkholyak$^{1}$,
        J. Stre\v{c}ka$^{2}$,
        M. Ja\v{s}\v{c}ur$^{2}$,
        and
        J. Richter$^3$}
\address{$^1$Institute for Condensed Matter Physics, NASU,\\
          1 Svientsitskii Street, L'viv-11, 79011, Ukraine\\
         $^2$Department of Theoretical Physics and Astrophysics,
          Institute of Physics, P. J. \v{S}af\'{a}rik University,
          Park Angelinum 9, 040 01 Ko\v{s}ice, Slovak Republic\\
         $^3$Institut f\"ur Theoretische Physik,
             Otto-von-Guericke-Universit\"at Magdeburg,
             P.O. Box 4120, 39016 Magdeburg, Germany
          }

\date{\today}

\begin{abstract}
The spin-1/2 $XXZ$ diamond chain is considered within the
Jordan-Wigner fermionization.
The fermionized Hamiltonian contains the interacting terms
which are treated within the Hartree-Fock approximation.
We obtain the ground-state magnetization curve of the model
for some particular cases and compare the results with
the exact diagonalization data for finite chains of 30 spins
and known exact results.
We also analyze the validity of the suggested approximation.
\end{abstract}



\maketitle

The spin-1/2 $XXZ$ diamond chain as a quantum frustrated system
is an interesting topic for the theoretical
research, since it exhibits many phenomena related to
interplay of quantum fluctuations and competing interactions.
The exact results for this model are available only for limited cases
\cite{tks,canova},
and the earlier studies used mainly the
numerical methods \cite{okamoto,okamoto2005,honecker}.
We suggest an analytical method based on the fermionization of the initial spin-1/2 model.
The present work is an extension of our previous study
of the $XX$ diamond chain \cite{xx} using
the Jordan-Wigner transformation and the Hartree-Fock approximation.
The goal of the paper is to explain
the properties of the frustrated model on the fermionic language
for a more general $XXZ$ model.
We accompany our approximate calculations
by the results of the exact diagonalization and also
compare them with the particular cases where the exact results are available
\cite{tks,canova}.

We consider the quantum spin-1/2 $XXZ$ model on the generally
distorted diamond chain (see Fig.1 in Ref.\onlinecite{xx}) with the following Hamiltonian:
\begin{eqnarray}
\label{ham-xxz-1}
H&{=}&\sum_{l=1}^N\left[
\sum_{\alpha=x,y,z}\left(
J_1^{\alpha} (s_{1,l}^\alpha s_{2,l}^\alpha {+} s_{3,l}^\alpha s_{1,l+1}^\alpha)
{+} J_2^{\alpha} s_{2,l}^\alpha s_{3,l}^\alpha
\right.\right.
\nonumber\\
&&\left.\left.
{+} J_3^{\alpha} (s_{1,l}^\alpha s_{3,l}^\alpha {+} s_{2,l}^\alpha s_{1,l+1}^\alpha)
\right)
-h\sum_{p=1}^3s_{p,l}^z \right].
\end{eqnarray}
Here, $s_{p,l}^{\alpha}$ ($\alpha=x,y,z$)
are the Pauli spin-1/2 operators with the first
index corresponding to a sublattice and the second index to a cell,
$J_p^x=J_p^y=J_p>0$, $J_p^z=\Delta J_p>0$,
$\Delta$ is the interaction anisotropy,
and $h$ is the external magnetic field (we set $g \mu_B = 1$).
Using the spin raising and lowering operators
$s_{p,l}^{\pm}=s_{p,l}^x \pm i s_{p,l}^y$
one can rewrite the $xy$-part of the Hamiltonian as a quadratic form
of the mentioned operators.
The $zz$ interaction,
due to the relation $s_{p,l}^z=s_{p,l}^+s_{p,l}^- -1/2$,
leads to the product
of four spin-lowering and raising operators.

Following the procedure described in Ref.\onlinecite{xx},
we specify the Jordan-Wigner transformation in the form:
$c_{1,l}=\prod_{p=1}^3\prod_{i=1}^{l-1} P_{q,i} s_{1,l}^-$,
$c_{2,l}=\prod_{p=1}^3\prod_{i=1}^{l-1} P_{q,i} P_{1,l}s_{1,l}^-$,
$c_{2,l}=\prod_{p=1}^3\prod_{i=1}^{l-1} P_{q,i} P_{1,l}P_{2,l}s_{1,l}^-$,
where $P_{q,l}=(-2s_{q,l}^z)=\exp(i\pi c_{q,l}^+ c_{q,l})$
is the Jordan-Wigner factor.
Here new operators $c_{p,l}$ satisfy the Fermi commutation relations.
The fermionic expression for the $z$-component of the spin operator
can be easily obtained as $s_{q,l}^z=c_{q,l}^+c_{q,l}-1/2$.
Therefore, the $zz$-part of the Hamiltonian contains
the four-fermion term.
The $xy$-part of the Hamiltonian in terms of new Fermi operators is as follows:
\begin{eqnarray}
\label{ham-xxz-2}
H_{xx}&{=}&\frac{1}{2}\sum_{l=1}^N\biggl[ (
J_1 (c_{1,l}^+ c_{2,l} +  c_{3,l}^+ c_{1,l+1})
+ J_2 c_{2,l}^+ c_{3,l}
\nonumber\\
&&+ J_3 (c_{1,l}^+P_{2,l} c_{3,l} +  c_{2,l}^+ P_{3,l} c_{1,l+1}) +
\mbox{h.c.}) \biggr].
\end{eqnarray}
Note, that the transformed Hamiltonian contains the fermion interaction for
terms proportional to the $J_3$ coupling which is usually assumed to be smaller than $J_2$.
We should also note the relation between the spins-1/2 and spinless fermions:
the spin-down (-up) state corresponds to the empty (filled) fermionic state;
the action of the fermion creation and annihilation operators
is analogous to the action of spin raising and lowering operators
in spin language.
The resulting Hamiltonian represents the interacting Fermi gas.
To proceed we use the Hartree-Fock approximation where all interacting terms
are factorized preserving all pair correlations between nearest neighbors of type
$\langle c_{p,l}^+ c_{q,m} \rangle$.
Thus, the Hamiltonian becomes a quadratic form in terms of Fermi operators.
It can be diagonalized using Fourier and Bogolyubov transformation,
and the thermodynamics and fermionic correlation functions are easily found.
However, it depends parametrically on the unknown contractions
$\langle c_{p,l}^+ c_{q,m} \rangle$ which have to be found self-consistently.

Solving the self-consistent equation we found that
the dimer-monomer ground state for the symmetric diamond chain \cite{tks}
is recovered.
Indeed, if $J_2\geq 2J_1$ and the external field $0<h<J_1\Delta+J_2(1+\Delta)/2$,
we obtain the following solutions for the elementary contractions:
$\langle c_{1,l}^+c_{1,l}\rangle=1$,
$\langle c_{2,l}^+c_{2,l}\rangle=\langle c_{3,l}^+c_{3,l}\rangle=1/2$,
$\langle c_{1,l}^+c_{2,l}\rangle=\langle c_{1,l}^+c_{3,l}\rangle=0$,
$\langle c_{2,l}^+c_{3,l}\rangle=-1/2$.
The ground state of the corresponding fermion model is as follows
$|GS\rangle=\prod_{l}c_{1,l}^+(c_{2,l}^+-c_{3,l}^+)/\sqrt{2}|0\rangle$
where $|0\rangle$ denotes the empty state.
Using the relation between fermionic and spin states
one can see that
it corresponds to the dimer-monomer state in spin language (see also Ref.\onlinecite{xx}).
\begin{figure}[t]
\begin{center}
\includegraphics[angle = 0, width = 0.69\linewidth]{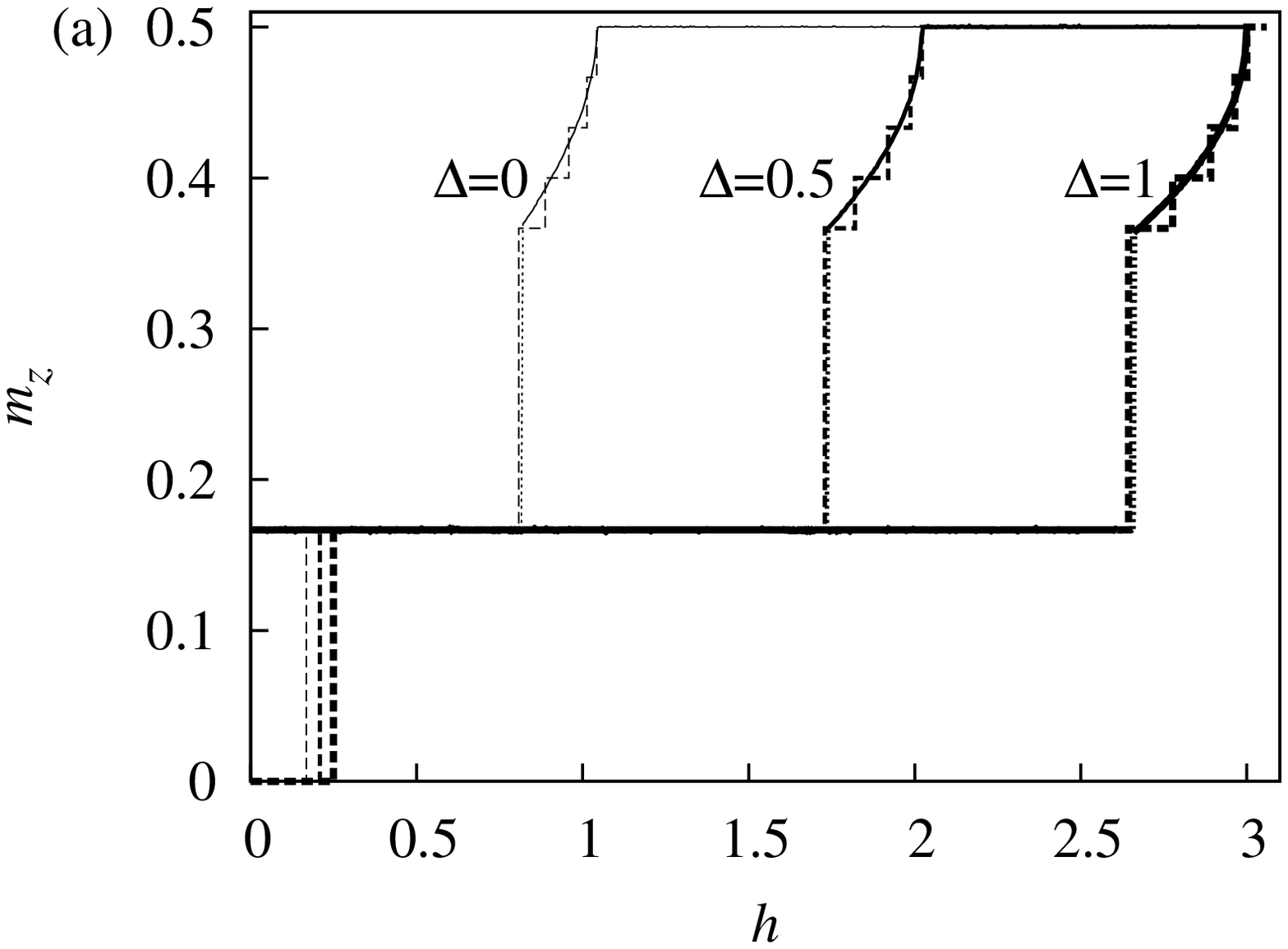}
\includegraphics[angle = 0, width = 0.69\linewidth]{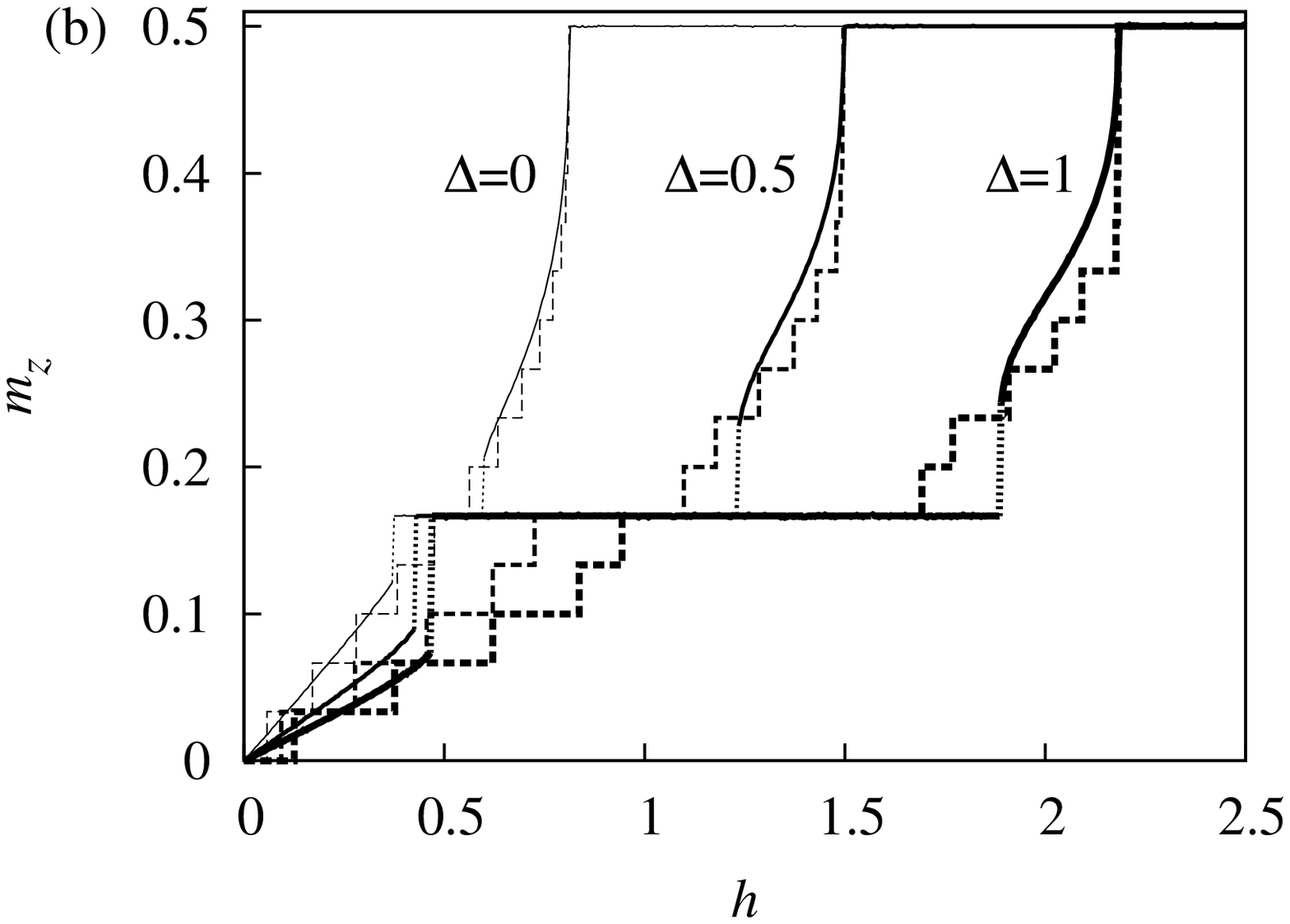}
\vspace*{-7mm}
\end{center}
\caption{The ground state magnetization versus external field
for
a) $J_1=J_3=1$, $J_2=1.75$;
b) $J_1=1$, $J_2=1.25$, $J_3=0.45$.
Step-like dashed lines represent the exact diagonalization data for 30 spins,
solid lines correspond to the approximated results, dotted lines indicate
the magnetization jumps.}
\label{fig_mz}
\end{figure}
The results for the symmetric diamond chain below the dimer-monomer limit
is shown in Fig.\ref{fig_mz}a.
We observe an excellent agreement with the exact diagonalization data
except for small fields
where our mean-field-type approximation produces the non-zero magnetization.
As an example of the distorted chain,
we have chosen the parameter set used previously for azurite \cite{kikuchi}.
In Fig.\ref{fig_mz}b we obtain a good agreement with the $XX$ limit.
Particularly, the magnetization curve shows the 1/3-plateau and zero-magnetization in zero field.
However, for the isotropic Heisenberg interaction the discrepancy between
the approximate and exact results raises up quickly below the upper critical field.
The exact diagonalization data for $\Delta=1$
seem to show a cusp in the magnetization curve
at 2/3 of the saturation magnetization.
It might be the sign that the magnetic cell is doubled  for this model.
As it was discussed in \cite{xx}, to describe this kind of behavior, it is necessary
to consider also non-uniform elementary contractions.
Another drawback of the Hartree-Fock approximation is
the artificial jumps of the magnetization.

We have also examined our method by comparison with the exact
results for the Ising-Heisenberg diamond chain obtained by means
of decoration-iteration procedure \cite{canova}. It is the special
case of the anisotropic diamond chain where the spins on the
vertical bond are coupled by the Heisenberg interaction whereas
all other couplings are of Ising type. The model (\ref{ham-xxz-1})
corresponds to the model considered in \cite{canova}, if we put
$J_1^{x,y}=J_3^{x,y}=0$, $J_2^{x,y}=\Delta J_2$,
$J_1^z=J_3^z=J_1$, $J_2^z=J_2$.
Depending on the ratio between the interaction
couplings, anisotropy and external field the system may stay in
different phases: $|{\rm
FRI}\rangle=\prod_{l}|\downarrow_{1,l}\rangle|\uparrow_{2,l}\uparrow_{3,l}\rangle$,
$|{\rm FRU}\rangle=\prod_{l}|\uparrow_{1,l}\rangle
(|\uparrow_{2,l}\downarrow_{3,l}\rangle-|\downarrow_{2,l}\uparrow_{3,l}\rangle)/\sqrt{2}$,
$|{\rm
SPP}\rangle=\prod_{l}|\uparrow_{1,l}\rangle|\uparrow_{2,l}\uparrow_{3,l}\rangle$.
Within our Hartree-Fock approach it is possible to recover the
ground state phase diagram and all phase boundaries. The mentioned
states in the fermion representation have the form: $|{\rm
FRI}\rangle=\prod_{l}c_{2,l}^+ c_{3,l}^+|0\rangle$, $|{\rm
FRU}\rangle=\prod_{l}c_{1,l}^+(c_{2,l}^+-c_{3,l}^+)/\sqrt{2}|0\rangle$,
$|{\rm SPP}\rangle=\prod_{l}c_{1,l}^+ c_{2,l}^+
c_{3,l}^+|0\rangle$.

To conclude, we have considered the approach based on the Jordan-Wigner fermionization
and subsequent Hartree-Fock approximation for the spin-1/2 anisotropic diamond chain.
We have revealed that such an approach recovers the exact results
for the phases characterized
by short-range correlations as, for instance, the dimer-monomer phase.
It also provides a good description of the ground state magnetization
for the symmetric diamond chain. However, for the distorted diamond chain
the inclusion of the $zz$ interactions between spins may lead to the qualitative change
of the magnetization curve which cannot be explained within the current approximate method.

\end{document}